\documentclass{PoS}
\usepackage{amsmath}
\usepackage{amssymb}
\usepackage{wrapfig}
\usepackage{graphicx}

\title{Monte Carlo studies of dynamical compactification of extra dimensions in a model of nonperturbative string theory 
        \footnote{Preprint number: KEK-TH-1860}}

\ShortTitle{Dynamical compactification of extra dimensions in the Euclidean IIB matrix model}

\author{Konstantinos N. Anagnostopoulos\\
        Physics Department, National Technical University of Athens
        Zografou Campus, 157-80 Zografou, Greece\\
        E-mail: \email{konstant@mail.ntua.gr}}

\author{\speaker{Takehiro Azuma}
        \thanks{T.A. thanks HPCI strategic program field 5 user support system for technical supports. 
The numerical simulations were carried out at KEKCC, NTUA het clusters, FX10 at Kyushu University and K computer through the HPCI System Research project (hp150082).
K.N.A.'s research is supported and implemented under the ARISTEIA II action  of the operational programme education and lifelong learning and is co-funded by the European Social Fund (ESF) and National Resources of Greece. The work of J.~N.\ was supported in part by Grant-in-Aid for Scientific Research (No.\ 23244057) from Japan Society for the Promotion of Science.}
        \\
        Institute for Fundamental Sciences, Setsunan University,
        17-8 Ikeda Nakamachi, Neyagawa,
        Osaka 572-8508, Japan\\
        E-mail: \email{azuma@mpg.setsunan.ac.jp}}
        
\author{Jun Nishimura\\
        High Energy Accelerator Research Organization (KEK) and Graduate University for Advanced
        Studies (SOKENDAI), 1-1 Oho, Tsukuba 305-0801, Japan\\
        E-mail: \email{jnishi@post.kek.jp}}        

\abstract{The IIB matrix model has been proposed as a non-perturbative definition of superstring theory. In this work, we study the Euclidean version of this model in which extra dimensions can be dynamically compactified if a scenario of spontaneously breaking the SO(10) rotational symmetry is realized. Monte Carlo calculations of the Euclidean IIB matrix model suffer from a very strong complex action problem due to the large fluctuations of the complex phase of the Pfaffian which appears after integrating out the fermions. We employ the factorization method in order to achieve effective sampling. We report on preliminary results that can be compared with previous studies of the rotational symmetry breakdown using the Gaussian expansion method.}

\FullConference{The 33rd International Symposium on Lattice Field Theory\\
		14 -18 July 2015\\
		Kobe International Conference Center, Kobe, Japan*}

\begin{document}
\section{Introduction}
Large-$N$ reduced models have been proposed as the non-perturbative definition of superstring theory. Especially, the IIB matrix model \cite{9612115} is one of the most successful proposals. The IIB matrix model is formally obtained by the dimensional reduction of ten-dimensional ${\cal N}=1$ super-Yang-Mills theory to zero dimensions. In the IIB matrix model, spacetime is
dynamically generated from the degrees of freedom of the bosonic matrices, despite the fact that it does not exist a priori in the model. Superstring theory is well-defined only in ten-dimensional spacetime, and it is an important question how our four-dimensional spacetime dynamically emerges. Monte Carlo studies of the IIB matrix model have a possibility to shed light on this question from a first principle calculation.

The Euclidean version of the IIB matrix model is obtained after a Wick rotation of the temporal direction. It has a manifest SO(10) rotational symmetry which, if spontaneously broken, yields
a spacetime compactified to lower dimensions. However, its numerical simulation has been hindered by the ``complex action problem'', because the Pfaffian obtained after integrating out the fermions is complex in general. 

Apart from the matrix models of superstring theory, there are many interesting systems that are plagued by the ``complex action problem''. Lattice gauge theories with a non-zero chemical potential are the ones that have attracted most of the attention in this context. In this work, we apply the ``factorization method'', which was originally proposed in ref. \cite{0108041} and generalized in ref. \cite{10094504}, to the Monte Carlo studies of the Euclidean version of the IIB matrix model. The IIB matrix model has also been studied analytically by the Gaussian Expansion Method (GEM) \cite{10070883,11081293}. Preliminary results of our Monte Carlo simulation are consistent with the GEM results and provide evidence that the factorization method is a successful approach to studying interesting systems that suffer from the complex action problem.
\section{Factorization method} \label{sec_fac}
Generally, it is difficult to numerically simulate the complex action system 
\begin{eqnarray}
 Z = \int dA e^{-S_0 + i \Gamma}. \label{complex_action}
\end{eqnarray}
Since $e^{-S_0 + i \Gamma}$ is not real positive, we cannot view it as a sampling probability in the Monte Carlo simulation. One way to calculate the vacuum expectation value (VEV) of an observable ${\cal O}$ is to use the reweighting $\langle {\cal O} \rangle = \frac{\langle {\cal O} e^{i \Gamma} \rangle_0}{\langle e^{i \Gamma} \rangle_0}$. Here, $\langle \cdots \rangle$ and $\langle \cdots \rangle_0$ are the VEV's for the original partition function $Z$ and the phase-quenched partition function $Z_0 = \int dA e^{-S_0}$, respectively. This is not an easy task since the phase $\Gamma$ may fluctuate wildly. In order to compute  $\langle
{\cal O} \rangle$ with given accuracy one needs O$(e^{\textrm{const.} V})$ configurations, where $V$ is the system size. This is called the ``sign problem'' or the ``complex action problem''.

Yet another problem is that the important configurations are different for different partition functions. This is called the ``overlap problem''. We are plagued with this overlap problem in trying to obtain the VEV $\langle {\cal O} \rangle$ through the simulation of the phase-quenched partition $Z_0$. 

The factorization method was proposed in order to reduce the overlap problem and achieve an importance sampling for the original partition function $Z$ \cite{0108041,10094504}. We select the set of the observables 
\begin{eqnarray}
 \Sigma = \{ {\cal O}_k | k=1,2,\cdots, n \}, \label{set_observables}
\end{eqnarray} 
which are strongly correlated with the phase $\Gamma$. In the following, we define the normalized observables ${\tilde {\cal O}}_k = {\cal O}_k/\langle {\cal O}_k \rangle_0$.
We employ the factorization property of the density of states $\rho (x_1, \cdots, x_n)$:
\begin{eqnarray}
 \rho (x_1, \cdots, x_n) = \langle \prod_{k=1}^n  \delta (x_k - {\tilde {\cal O}}_k) \rangle = \frac{1}{C} \rho^{(0)} (x_1, \cdots, x_n) w(x_1, \cdots, x_n). \label{factorization}
\end{eqnarray}
The constant $C = \langle e^{i \Gamma} \rangle_0$ is irrelevant in the following. $\rho^{(0)} (x_1, \cdots, x_n) = \langle \prod_{k=1}^n  \delta (x_k - {\tilde {\cal O}}_k) \rangle_0$ is the density of states in the phase-quenched model. $w(x_1, \cdots x_n) = \langle e^{i \Gamma} \rangle_{x}$ is the VEV in the constrained system
\begin{eqnarray}
 Z_{x} = \int dA e^{-S_0} \prod_{k=1}^n  \delta (x_k - {\tilde {\cal O}}_k). \label{factorization2}
\end{eqnarray}
When the system size $V$ goes to infinity, the VEV's are given by $\langle {\tilde {\cal O}}_k \rangle = {\bar x}_k$, where $({\bar x}_1, \cdots, {\bar x}_n)$ is the position of the peak of $\rho (x_1, \cdots, x_n)$. This can be obtained by solving the saddle-point equation
\begin{eqnarray}
 \lim_{V \to +\infty} \frac{\partial}{\partial x_k} \left\{ \frac{1}{V} \log \rho^{(0)} (x_1, \cdots, x_n) \right\}= -  \frac{\partial}{\partial x_k} \left\{ \lim_{V\to +\infty} \frac{1}{V} \log w(x_1, \cdots, x_n) \right\}. \label{saddle_point}
\end{eqnarray}
When we properly choose the maximal set of the observables $\Sigma$, we achieve effective importance sampling for the original partition function $Z$ \cite{10094504}.
\section{Euclidean version of the IIB matrix model} \label{sec_IKKT}
We study the IIB matrix model \cite{9612115}, which is defined by the following partition function:
\begin{eqnarray}
 Z = \int dA d\psi e^{-(S_{\textrm{b}} + S_{\textrm{f}})}, \label{IKKT_partition}
\end{eqnarray}
where the bosonic part $S_{\textrm{b}}$ and the fermionic part $S_{\textrm{f}}$ are respectively 
\begin{eqnarray}
 S_{\textrm{b}} &=& - \frac{1}{4g^2} \textrm{tr} [A_{\mu}, A_{\nu}]^{2}, \label{IKKT_boson} \\
 S_{\textrm{f}} &=& \frac{1}{2g^2} \textrm{tr} \left( \psi_{\alpha} ({\cal C} \Gamma_{\mu})_{\alpha \beta} [A_{\mu}, \psi_{\beta}] \right). \label{IKKT_fermion}
\end{eqnarray}
The bosons $A_{\mu}$ ($\mu =1,2, \cdots, 10$) and the Majorana-Weyl spinors $\psi_{\alpha}$ ($\alpha =1,2, \cdots, 16$) are  $N \times N$ traceless hermitian matrices.  In the following, without loss of generality we set $g^2 N=1$. The indices are contracted by the Euclidean metric after the Wick rotation. $\Gamma_{\mu}$ are the $16 \times 16$ Gamma matrices after the Weyl projection, and ${\cal C}$ is the charge conjugation matrix. 
This model has the SO(10) rotational symmetry. In ref. \cite{9803117}, it is shown that the partition function is positive definite without cutoffs.

This model is formally obtained by the dimensional reduction of ten-dimensional ${\cal N}=1$ super Yang-Mills theory to zero dimensions. The IIB matrix model has the ${\cal N}=2$ supersymmetry
\begin{eqnarray}
 \delta^{(1)}_{\varepsilon} A_{\mu} = i \varepsilon ({\cal C} \Gamma_{\mu}) \psi, \ \ \delta^{(1)}_{\varepsilon} \psi = \frac{i}{2} [A_{\mu}, A_{\nu}] \Gamma^{\mu \nu} \varepsilon, \ \ \delta^{(2)}_{\varepsilon} A_{\mu} = 0, \ \ \delta^{(2)}_{\varepsilon} \psi = \varepsilon. \label{IKKT_SUSY}
\end{eqnarray}
For the linear combination ${\tilde \delta}^{(1)}_{\varepsilon} = \delta^{(1)}_{\varepsilon} + \delta^{(2)}_{\varepsilon}$ and  ${\tilde \delta}^{(2)}_{\varepsilon} = i (\delta^{(1)}_{\varepsilon} - \delta^{(2)}_{\varepsilon})$, we have 
\begin{eqnarray}
 [{\tilde \delta}^{(a)}_{\varepsilon}, {\tilde \delta}^{(b)}_{\xi}] A_{\mu} = -2i \delta^{ab} \varepsilon ({\cal C} \Gamma_{\mu}) \xi, \ \ [{\tilde \delta}^{(a)}_{\varepsilon}, {\tilde \delta}^{(b)}_{\xi}] \psi = 0, \ \ (a,b=1,2). \label{IKKT_SUSY2}
\end{eqnarray}
This leads to the interpretation of the eigenvalues of the bosonic matrices $A_{\mu}$ as the spacetime coordinates. 
Hence, the spontaneous symmetry breakdown (SSB) of the SO(10) rotational symmetry is identified with the dynamical compactification of the extra dimensions.

The order parameters of the SSB of the SO(10) rotational symmetry are the eigenvalues $\lambda_{n}$ ($n=1,2,\cdots,10$) of the ``moment of inertia tensor''
\begin{eqnarray}
 T_{\mu \nu} = \frac{1}{N} \textrm{tr} (A_{\mu} A_{\nu}), \label{Tmunu}
\end{eqnarray}
which are ordered as $\lambda_{1} > \lambda_{2} > \cdots > \lambda_{10}$ before taking the expectation value. If $\langle \lambda_{1} \rangle, \cdots, \langle \lambda_{d} \rangle$ grow and $\langle \lambda_{d+1} \rangle, \cdots, \langle \lambda_{10} \rangle$ shrink in the large-$N$ limit, this suggests the SSB of the SO(10) rotational symmetry to SO$(d)$ and hence the dynamical compactification of ten-dimensional spacetime to $d$ dimensions. This scenario has been studied via GEM in ref. \cite{11081293}. The results of the studies of the SO$(d)$ symmetric vacua for $2 \leq d \leq 7$ are summarized as follows:
\begin{enumerate}
\item{The extent of the shrunken directions $r = \lim_{N\to \infty} \sqrt{\lambda_n}$ ($n=d+1, \cdots,10$) is $r^2 \simeq 0.155$, which does not depend on $d$ (universal compactification scale).}
\item{The ten-dimensional volume of the Euclidean spacetime does not depend on $d$ except $d=2$ (constant volume property). For the extent of the extended directions $R = \lim_{N \to \infty} \sqrt{\lambda_n}$ ($n=1,2,\cdots,d$), the volume is $V = R^d r^{10-d} = l^{10}$, with $l^2 \simeq 0.383$.}
\item{The free energy takes the minimum value at $d=3$, which suggests the dynamical emergence of {\it three}-dimensional spacetime.}
\end{enumerate}
In ref. \cite{10070883}, the six-dimensional version of the Euclidean IIB matrix model was studied via GEM, and the six-dimensional version also turns out to have these three properties. The same model was studied numerically in ref. \cite{13066135}, and the results are consistent with the GEM results.


\begin{wrapfigure}{r}{65mm}
\vspace*{-10mm}
 \begin{flushright}
 \hspace*{-2mm} \scalebox{0.5750}{\includegraphics{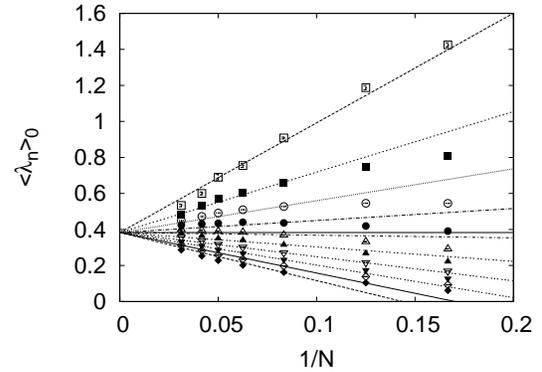}}
 \end{flushright}
 \vspace*{-6mm}
\caption{The VEV $\langle \lambda_n \rangle_0$ with respect to the phase-quenched partition fucntion $Z_0$ up to $N=32$.}
\label{pq_result} 
\end{wrapfigure}
Next, we review the mechanism of the dynamical compactification of spacetime in the Euclidean IIB matrix model \cite{0003223}. Integrating out the fermions, we have
\begin{eqnarray}
 \int d \psi e^{-S_{\textrm{f}}} = \textrm{Pf} {\cal M}, \label{pfaffian_M}
\end{eqnarray}
where ${\cal M}_{a\alpha, b \beta} = -i f_{abc} ({\cal C} \Gamma_{\mu})_{\alpha \beta} A_{\mu}^c$ is a $16 (N^2-1) \times 16(N^2-1)$ anti-symmetric matrix. The indices $a,b,c$ run over $1,2,\cdots,N^2-1$, and $f_{abc}$ are the structure constants of SU$(N)$. $A_{\mu}^c$ are the coefficients in the expansion $A_{\mu} = \sum_{c=1}^{N^2-1} A_{\mu}^c T^c$ with respect to the SU$(N)$ generators $T^c$. Under the transformation $A_{10} \to - A_{10}$, $\textrm{Pf} {\cal M}$ becomes complex conjugate. We define the phase of the Pfaffian $\Gamma$ as $\textrm{Pf} {\cal M} = |\textrm{Pf} {\cal M}| e^{i \Gamma}$. $\textrm{Pf} {\cal M}$ is real for the nine-dimensional configuration $A_{10}=0$. When the configuration is $d$-dimensional ($3 \leq d < 9$), we find  $\frac{\partial^m \Gamma}{\partial A_{\mu_1}^{a_1} \cdots \partial A_{\mu_m}^{a_m}} = 0$
for $m=1,2,\cdots,9-d$, because the configuration is at most nine-dimensional up to the $(9-d)$-th order of the perturbations. Thus, the phase of $\textrm{Pf} {\cal M}$ becomes more stationary for the lower dimensions. The numerical results in ref. \cite{0003208} also suggest that there is no SSB of the rotational symmetry in the phase-quenched model. We calculate $\langle \lambda_n \rangle_0$ numerically, where $\langle \cdots \rangle_0$ is the VEV with respect to the phase-quenched partition function
\begin{eqnarray}
 Z_0 = \int d A e^{-S_{\textrm{b}}} |\textrm{Pf} {\cal M}|. \label{pq_partition}
\end{eqnarray}
We use the Rational Hybrid Monte Carlo (RHMC) algorithm, whose details are presented in Appendix A of ref. \cite{13066135}. The result in fig. \ref{pq_result} shows that $\langle \lambda_n \rangle_0$ converge to $l^2 \simeq 0.383$ at large $N$ for all $n=1,2,\cdots,10$. This suggests that there is no SSB of the SO(10) rotational symmetry, and that the result is consistent with the constant volume property.
\section{Results} \label{sec_results}
The model (\ref{IKKT_partition}) suffers from a strong complex action problem, and we apply the factorization method to this system. It turns out to be sufficient to constrain only one eigenvalue $\lambda_{n}$; namely the choice of the set $\Sigma$ in eq. (\ref{set_observables}) should be $\Sigma = \left\{ \lambda_n \right\}$. This is because the larger eigenvalues do not affect much the fluctuation of the phase. This choice of $\Sigma$ is similar to that of the six-dimensional version of the IIB matrix model \cite{13066135}. When we constrain $\lambda_{n}$, the eigenvalues $\lambda_{n}, \lambda_{n+1}, \cdots \lambda_{10}$ take the small value, which corresponds to the SO$(d)$ symmetric vacuum, with $n = d+1$. This leads us to simulate the partition function of the constrained system
\begin{eqnarray}
 Z_{n,x} = \int dA e^{-S_{\textrm{b}}} |\textrm{Pf} {\cal M}| \delta (x - {\tilde \lambda}_n), \label{nx-factorization}
\end{eqnarray}
which is simulated via the RHMC algorithm. The ratio ${\tilde \lambda}_n = \lambda_n/ \langle \lambda_n \rangle_0$ corresponds to the square of the ratio of the extents of the extended and shrunken directions $(r/l)^2$, in the SO$(d)$ vacua with $n=d+1$. The saddle-point equation (\ref{saddle_point}) is now simplified as
\begin{eqnarray}
 & & \frac{1}{N^2} f^{(0)}_n (x) = - \frac{d}{dx}\frac{1}{N^2} \log w_n(x), \textrm{ where } \label{saddle_point2} \\
 & & f^{(0)}_n (x) = \frac{d}{dx} \log \langle \delta (x - {\tilde \lambda}_n ) \rangle_0, \ \ w_n (x) = \langle e^{i \Gamma} \rangle_{n,x} = \langle \cos \Gamma \rangle_{n,x}, \label{saddle_point3} 
\end{eqnarray}
in the large-$N$ limit. $\langle \cdots \rangle_{n,x}$ is the VEV of the partition function $Z_{n,x}$. We have $\langle e^{i \Gamma} \rangle_{n,x} = \langle \cos \Gamma \rangle_{n,x}$, because under the transformation $A_{10} \to - A_{10}$ the Pfaffian $\textrm{Pf} {\cal M}$ becomes complex conjugate while the bosonic action (\ref{IKKT_boson}) and the eigenvalues of the tensor (\ref{Tmunu}) are invariant.
The solution of the saddle-point equation (\ref{saddle_point2}) ${\bar x}_n$ gives the VEV $\langle {\tilde \lambda}_n \rangle = {\bar x}_n$ in the SO$(d)$ vacuum with $n=d+1$. Solving this saddle-point equation amounts to finding the minimum of the free energy
\begin{eqnarray}
 {\cal F}_{\textrm{SO}(d)} (x) = - \frac{1}{N^2} \log \rho_n (x), \textrm{ where } \rho_n (x) = \langle \delta (x - {\tilde \lambda}_{n}) \rangle, \label{free_energy_sod}
\end{eqnarray}
in the SO$(d)$ vacuum with $n=d+1$. The GEM result suggests that the free energy takes the minimum for the SO(3) vacuum. In order to reduce the CPU costs, we focus on the $n=3,4,5$ cases, which correspond to the SO(2), SO(3), SO(4) vacua, respectively. 

\begin{figure}[htbp]
\vspace*{-4mm}
 \begin{center}
 \hspace*{+0mm} \scalebox{0.575}{\includegraphics{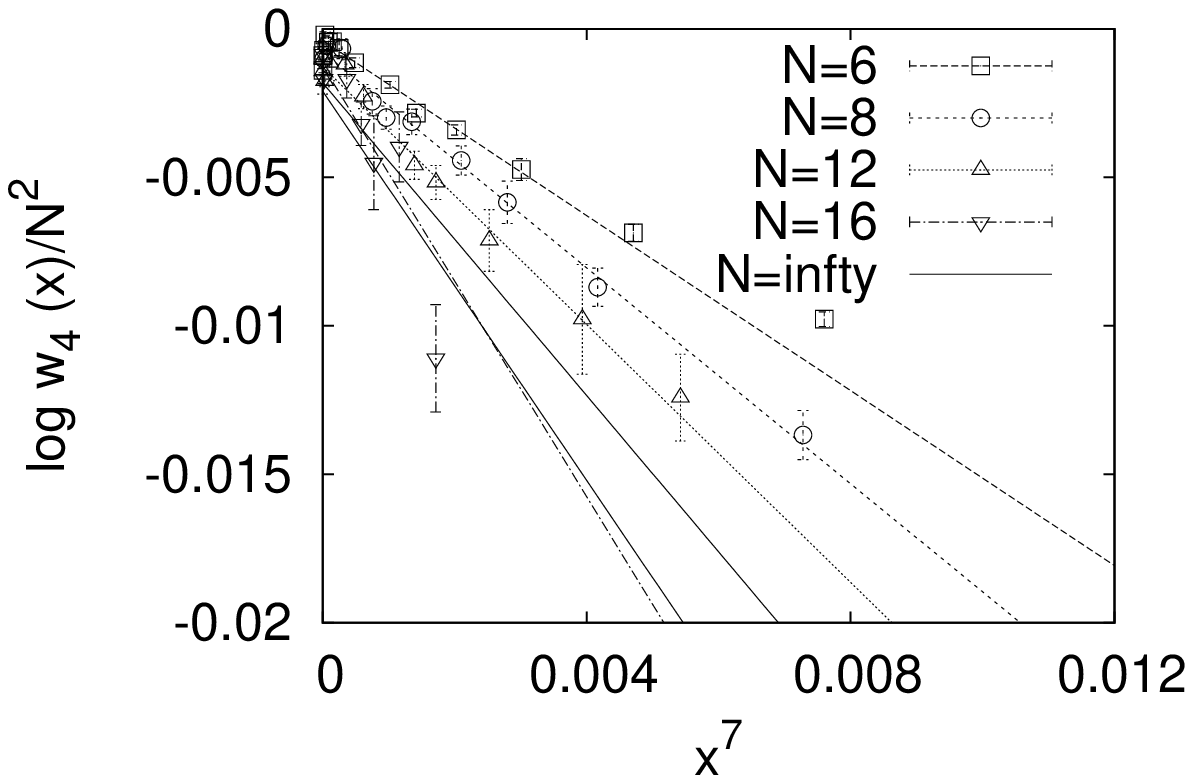}\includegraphics{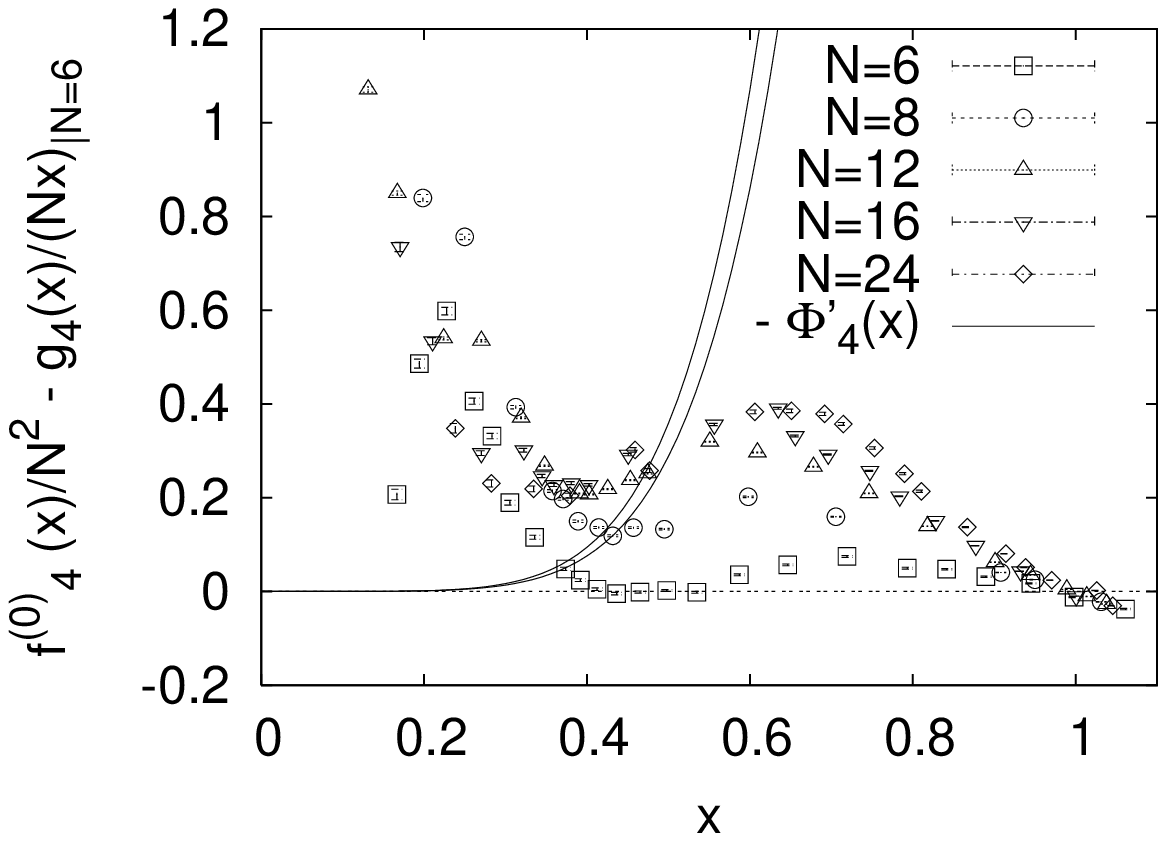}}
 \end{center}
 \vspace*{-6mm}
\caption{(LEFT) $\log w_{4} (x)$ against $x^{7}$ at small $x$. (RIGHT) $\frac{1}{N^2} f^{(0)}_4 (x) - \frac{g_4 (x)}{Nx}$ up to $N=24$. Its intersection with $- \frac{d}{dx} \Phi_n (x)$ gives the solution ${\bar x}_4$.}
\label{factorization_result} 
\end{figure}

In fig. \ref{factorization_result} (LEFT) we plot $\log w_n(x)$ for $n=4$ up to $N=16$, where we observe a good scaling behavior at small $x$
\begin{eqnarray}
 \frac{1}{N^2} \log w_n (x) \simeq - a_{n} x^{11-n} - b_{n}. \label{scale-w}
\end{eqnarray}
The coefficients $a_{n}$ and $b_{n}$ are obtained for each $N$, by fitting the data. Then, we extrapolate the coefficients $a_{n}, b_{n}$ and obtain the large-$N$ limit, which corresponds to $\Phi_n (x) = \lim_{N \to +\infty} \frac{1}{N^2} \log w_n (x)$. This is represented by the solid line in fig. \ref{factorization_result} (LEFT).

The function $f^{(0)} (x)$ has a scaling behavior around $0.4 \leq x \leq 1$
\begin{eqnarray}
 \frac{x}{N} f^{(0)}_n (x) \simeq g_n (x), \textrm{ where } g_n (x) = c_n (x-1) + d_n (x-1)^2. \label{f0_scale1}
\end{eqnarray}
Subtracting this effect in order to reduce finite-$N$ effects, we plot $\frac{1}{N^2} f^{(0)}_n (x) - \frac{g_n(x)}{Nx}$ for $n=4$ in fig. \ref{factorization_result} (RIGHT). We find that the results scale reasonably well up to $N=24$ in the small-$x$ region $x \leq 0.4$. This implies the hard-core potential structure at small $x$. In the six-dimensional version of the IIB matrix model, this effect is absent in the one-loop approximation \cite{0108041}, but is observed in the full model without one-loop approximation \cite{13066135}. The intersection of $\frac{1}{N^2} f^{(0)}_n (x) - \frac{g_n(x)}{Nx}$ and $- \frac{d}{dx} \Phi_n (x)$ represents the solution of the saddle-point equation (\ref{saddle_point2}). Fig. \ref{factorization_result} (RIGHT) shows that the solution ${\bar x}_n$ is close to $\frac{r^2}{l^2} \simeq \frac{0.155}{0.383} = 0.404 \cdots$ for $n=4$. For $n=3,5$, too, we have obtained similar results, and the solution ${\bar x}_n$ is close to 0.404. This is consistent with the ``universal compactification scale'' property.

\begin{wrapfigure}{r}{65mm}
\vspace*{-10mm}
 \begin{flushright}
 \hspace*{-2mm}  \scalebox{0.575}{\includegraphics{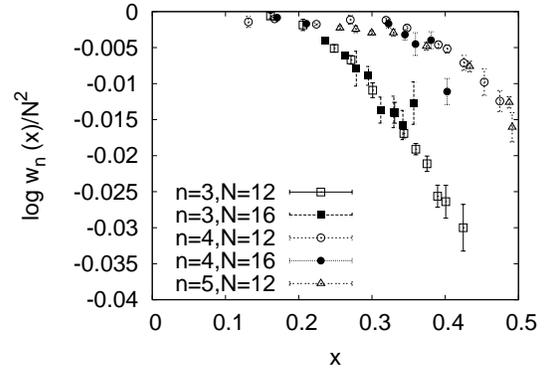}}
 \end{flushright}
 \vspace*{-6mm}
\caption{$\frac{1}{N^2} \log w_n (x)$ for $n=3,4$, $N=12,16$ and $n=5$, $N=12$.}
\label{free_result} 
\end{wrapfigure}

Next, we compare the free energy (\ref{free_energy_sod}) for the SO$(d)$ vacuum. The free energy at $x = {\bar x}_n$ is 
\begin{eqnarray}
 \hspace*{-3mm} {\cal F}_{\textrm{SO}(d)} = \int^1_{{\bar x}_n} dx \frac{1}{N^2} f^{(0)}_n (x) - \frac{1}{N^2} \log w_n ({\bar x}_n), \label{free_energy_sod2}
\end{eqnarray}
with $n=d+1$. Due to the scaling behavior (\ref{f0_scale1}), the first term of the r.h.s of eq. (\ref{free_energy_sod2}) vanishes at large $N$. Thus we compare $\frac{1}{N^2} \log  w_n ({\bar x}_n)$. From fig. \ref{free_result}, we see that the free energy ${\cal F}_{\textrm{SO}(2)}$ is much higher than ${\cal F}_{\textrm{SO}(3)}$ and ${\cal F}_{\textrm{SO}(4)}$ around $x \simeq 0.4$. It is still difficult to determine whether the SO(3) or the SO(4) vacuum is energetically favored. More analysis will be reported elsewhere.

\section{Conclusion} \label{sec_conclu}
In this work, we have performed Monte Carlo simulations of the Euclidean version of the IIB matrix model using the factorization method, in order to study the dynamical compactification of the extra dimensions.  The results turn out to be consistent with the GEM predictions. We have seen that in the phase-quenched model there is no SSB of the SO(10) rotational symmetry, and that the volume of spacetime is consistent with the GEM results. The function $f^{(0)}_n (x)$ has a hard-core potential structure, and as a result of that, the computed shrunken dimensions are found to be consistent with the GEM results. Also, we have succeeded in finding that the SO(2) vacuum is energetically disfavored, compared to the SO(3) or SO(4) vacuum. The results of the Lorentzian version of the IIB matrix model, where (3+1)-dimensional spacetime is found to expand dynamically \cite{11081540}, and the scenario discussed in this work, suggest that the physical interpretation of the Euclidean IIB matrix model needs to be further investigated.



\end{document}